# ***ECONOMICS UNCHAINED: INVESTIGATING THE ROLE OF CRYPTOCURRENCY, BLOCKCHAIN AND INTRICACIES OF BITCOIN PRICE FLUCTUATIONS***

ISHMEET MATHAROO

TERI SCHOOL OF ADVANCED STUDIES

OCTOBER,2023

## ***ABSTRACT***

This research paper presents a thorough economic analysis of Bitcoin and its impact. We delve into fundamental principles, and technological evolution into a prominent decentralized digital currency. Analysing Bitcoin's economic dynamics, we explore aspects such as transaction volume, market capitalization, mining activities, and macro trends. Moreover, we investigate Bitcoin's role in economy ecosystem, considering its implications on traditional financial systems, monetary policies, and financial inclusivity. We utilize statistical and analytical tools to assess equilibrium , market behaviour, and economic . Insights from this analysis provide a comprehensive understanding of Bitcoin's economic significance and its transformative potential in shaping the future of global finance. This research contributes to informed decision-making for individuals, institutions, and policymakers navigating the evolving landscape of decentralized finance.

## ***INTRODUCTION***

Cryptocurrency is a digital currency that is used for the medium of exchange. As the financial world continues its transformation, cryptocurrencies redefined traditional monitory system. The genesis of Bitcoin in 2009 marked the starting of a transformative era, cryptocurrency introduced a peer-to-peer electronic cash system that operates beyond the traditional institutes of finances. The rise of cryptocurrencies has opened doors for economics sovereignty to access the financial services. Cryptocurrency introduced a decentralised monetary system that challenged the traditional banking system and central authorities. This decentralisation has the power to grant access to economic empowerment and finance. As cryptocurrencies are not issued by central authorities therefore, their value is mostly obtained from the scale of participation in market. Cryptocurrency owners store their asset in a digital system of decentralised network. The rise of cryptocurrency has led to a dynamic shift in financial landscapes, attention from individual, businesses, investors and government. However, cryptocurrency ecosystem faces an array of challenges. It's uncertainties, security concerns, market volatility and financial obstacles to widespread adoption harnessed the benefits of cryptocurrency

## ***BITCOIN: A BRIEF REVIEW***

Bitcoin, often known as "digital gold", born amidst the global financial crises of 2008.The identity of ***Satoshi Nakamoto***, creator of bitcoin remains unknown. ***Nakamoto*** published the bitcoin and began developing the software in 2008, eventually launching it in 2009.Since the creation of bitcoin in 2009, a lot more cryptocurrencies have been introduced. Bitcoin is the most successful one. It is a decentralised digital currency without a central bank, that can be sent on a peer-to-peer bitcoin network without the need of intermediaries. Transactions are verified by network nodes through cryptography and recorded in a pubic distributed ledger called a Blockchain. Features of bitcoin are as follows; (source: ***KarlStrom) -***

-- The money supply is controlled by an algorithm, the workings of which are in the public domain which then in depended on central bank monetary policy.

-- Its transactions are decentralised and non-hierarchical.

-- It serves as anonymity as its electronic wallets are not directly connected to their respective owners.

This research paper embarks comprehensive study of Bitcoin, analysis its economic impact and operations. It aims to provide a deep dive into the world of cryptocurrencies by examining the technological foundations, regulatory landscapes, adoption of bitcoins, blockchain technology. Through careful analysis and evaluation, we aim to contribute a conclusion on the cryptocurrencies in this rapidly evolving global financial landscape.

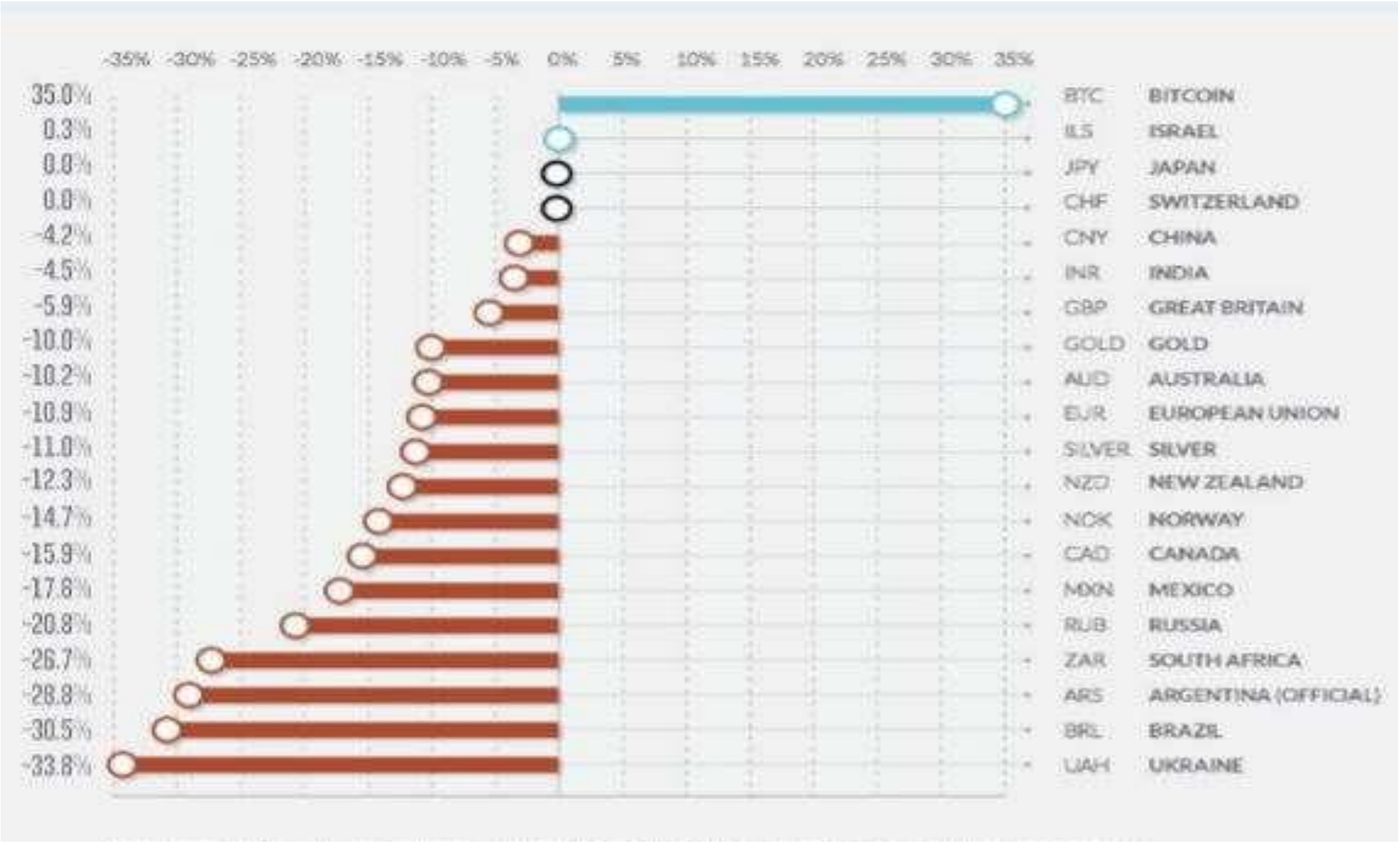

## ***MECHANISM***

Bitcoin operates on a mechanism known as Blockchain which is a decentralised distributive leger. It uses a proof of work consensus algorithm to validate and secure transaction. The blockchain consist of blocks. Each block contain transaction. Here miners compete to add new block to the chain and in return they are rewarded with newly created bitcoin and transaction fees.

***BLOCKCHAINS-***

Bitcoin operates on blockchain, it contains all transaction data from anyone who uses bitcoin. Transactions are contained into block each block is linked to the previous block forming a chain of bocks. Blocks are created approximately every 10 minutes on average. The total size of bitcoin blockchain in September 2021 was over 335 GB. By September 2021, the bitcoin blockchain add over 700000 blocks. The maximum block size for bitcoin is 1 MB.

***TRANSACTIONS-***

Transaction is cryptographically signed to ensure its authenticity. It involves the transfer of bitcoin value between bitcoin addresses. Each transaction have sources of fund and destination of fund. Later, transactions are grouped together and added to a mempool. As per the recent data, a bitcoin network can process an average of 7 transactions per second. Bitcoin blockchain has surpasses 700 million number of transactions by September 2021.

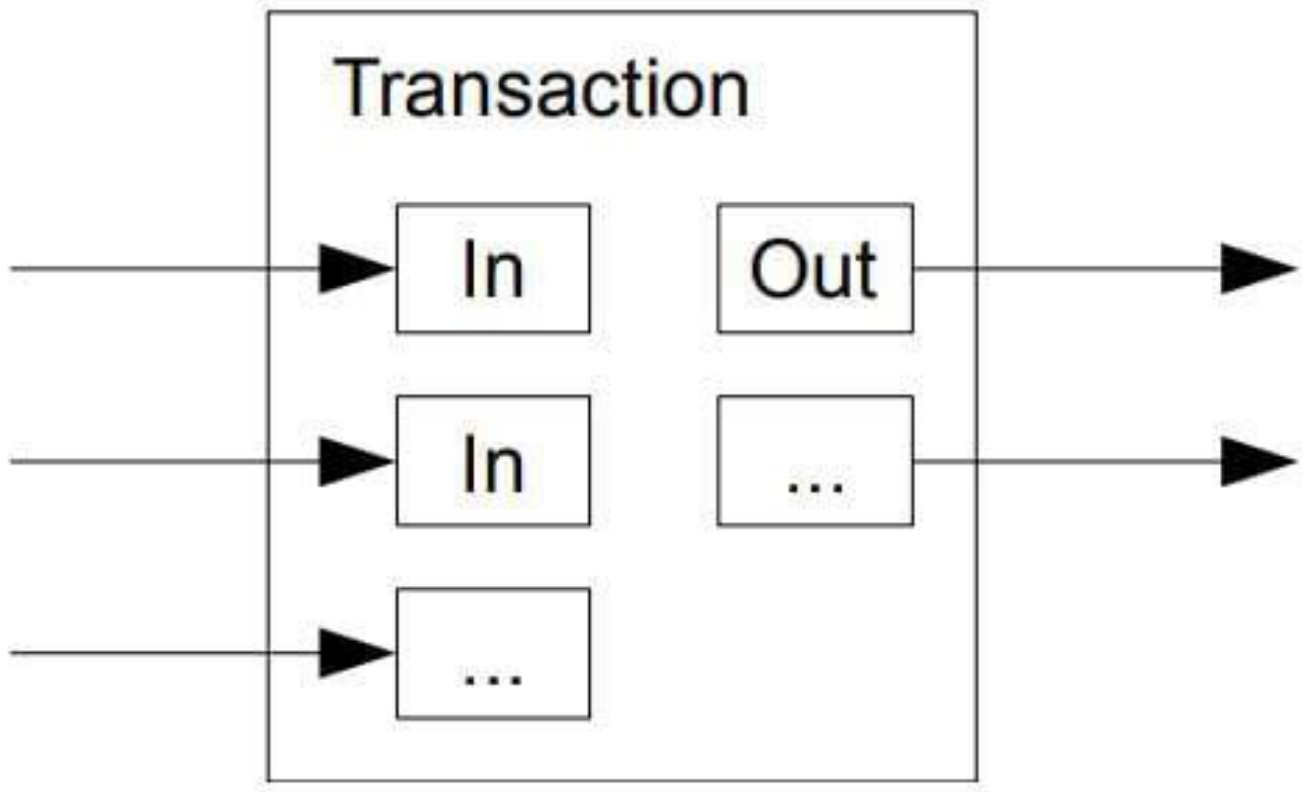

***PEER-TO-PEER NETWORK-***

To validate and record transaction, bitcoin operated on a decentralised network of nodes. This network of decentralisation ensures no single individual or authority controls the system.

***CONSENSUS MEACHANISM-***

To validate or secure transaction on bitcoin network PoW-Proof of Work is the consensus algorithm.

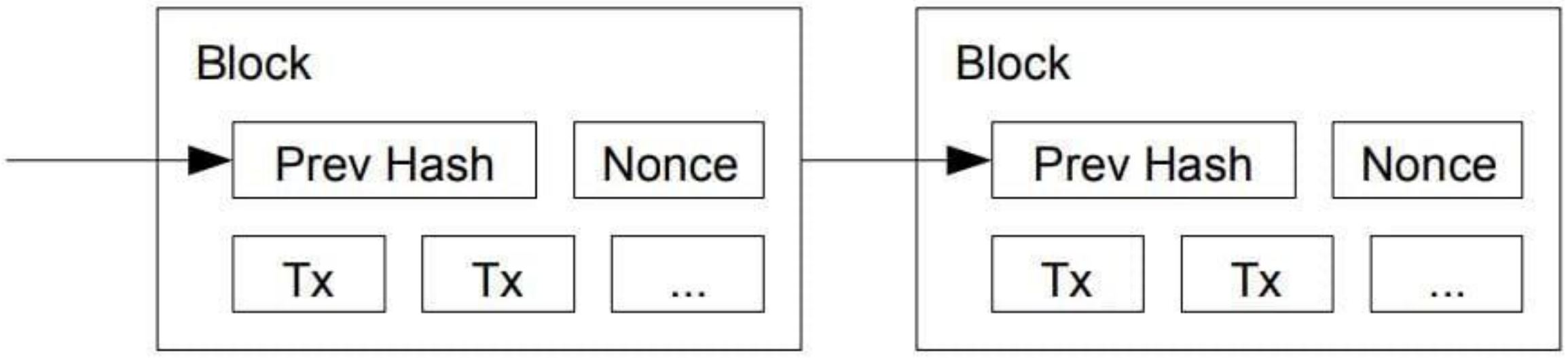

***MINING-***

PoW calculations are performed using specialised computer hardware which involves mining. Miner uses specialised hardware known as ASIC [Application Specific Integrated Circuit) for solving the proof of work algorithm used in bitcoin. Mining bitcoin is a competitive process. It plays a critical role in validating transaction and maintaining the decentralised nature of bitcoin blockchain. The difficulty of mining algorithm adjusts approximately every 2 weeks to maintain the target block time of 10 minutes.

***MINING REWARD-***

Nearly every four years, the reward of mining or new block id halved. This process helps to reduce the rate at which new bitcoins are created, following a predictable supply scheduled

***BLOCKCHAIN SUPPLY-***

By September 2021, around 18.8 million bitcoins had been mined out of the maximum limit of 21 million. The maximum supply limit of 21million makes it a deflationary asset. This scarcity helps to maintain its value.

# ***TRADING***

Trading of Bitcoin involves analysing various economic and market factors to make informed decisions about buying, selling, or holding the cryptocurrency. Economic analysis in regard of Bitcoin often includes assessing macroeconomic trends, market sentiment, regulatory developments, supply-demand dynamics. The key aspects to consider when economically analysing Bitcoin are as follows

1. **Macro-Economic Factors:**
    - **Inflation and Monetary Policy:** Bitcoin is often considered a hedge against inflation due to its capped supply around 21 million coins. Monitoring global inflation rates and central bank monetary policies is crucial to be understood for the trading of bitcoin .

- **Interest Rates:** Changes in interest rates can influence investment behaviour and affect the attractiveness of Bitcoin compared to traditional assets like bonds or savings accounts.

2. **Market Analysis:**

- **Technical Analysis:** Analysing price charts, patterns, volume, and other trading indicators can help identify trends and entry and exit points.

3. **Supply-Demand Dynamics:**

- **Bitcoin Halving Events:** Understand the impact of halving events, which reduce the rate of new Bitcoin creation and may influence its price due to reduced supply growth.
- **Market Liquidity:** Consider trading volume, order book depth, and liquidity levels
- to assess the ease of buying or selling Bitcoin.

4. **Global Economic Events:**

- **Geopolitical Tensions:** Political instability, trade disputes, or other geopolitical events can affect Bitcoin's price as investors seek alternative assets.
- **Market Crashes:** Assess the correlation between traditional markets and Bitcoin during economic downturns is a major event to keep an eye on.

5. **Mining Activity**:

- Monitor hash rate and mining difficulty as indicators of the network's health and security

Moreover, Bitcoin trading involves inherent risks, and no analysis can guarantee profits.

## ***EQUILIBRIUM***

In the regard to the Bitcoin market, reaching an equilibrium involves a balance between the supply of Bitcoin determined by mining and circulation and the demand for Bitcoin determined by investor interest, adoption, and usage.

**Supply of Bitcoin:**

Bitcoin is created through a process called mining, where miners use computational power to validate transactions and add them to the blockchain. This process also introduces new bitcoins into the system as rewards for miners, the reward for mining new blocks is halved, reducing the rate at which new bitcoins are introduced into the system. This halving event is designed to gradually decrease the supply of new bitcoins over time until the maximum supply of 21 million bitcoins is reached.

**Demand for Bitcoin:**

Demand for Bitcoin is driven by investors seeking to diversify their portfolios, hedge against inflation, or speculate on its price movements. As more merchants accept Bitcoin as a form of payment and more users engage with the cryptocurrency for transactions or investment, the demand for Bitcoin increases.

**Price Determination:**

The price of Bitcoin is determined by the interplay of supply and demand in the market. When demand exceeds supply, prices tend to rise, and when supply exceeds demand, prices tend to fall. The market constantly adjusts to new information, news, and events, influencing trader behaviour and shaping price movements.

**Market Dynamics:**

Bitcoin is known for its price volatility due to factors such as speculative trading, news events, regulatory developments, and macroeconomic shifts. This volatility can sometimes delay the attainment of a stable equilibrium.

**Long-Term Equilibrium:**

Over the long term, as Bitcoin adoption increases and its supply growth slows due to halving events, a more stable equilibrium may be achieved where supply and demand reach a more balanced state. It's important to note that achieving a stable equilibrium in the Bitcoin market is a dynamic process, influenced by a multitude of factors, include. Consequently, the Bitcoin market can experience periods of both stability and volatility.

(source : coingeek )

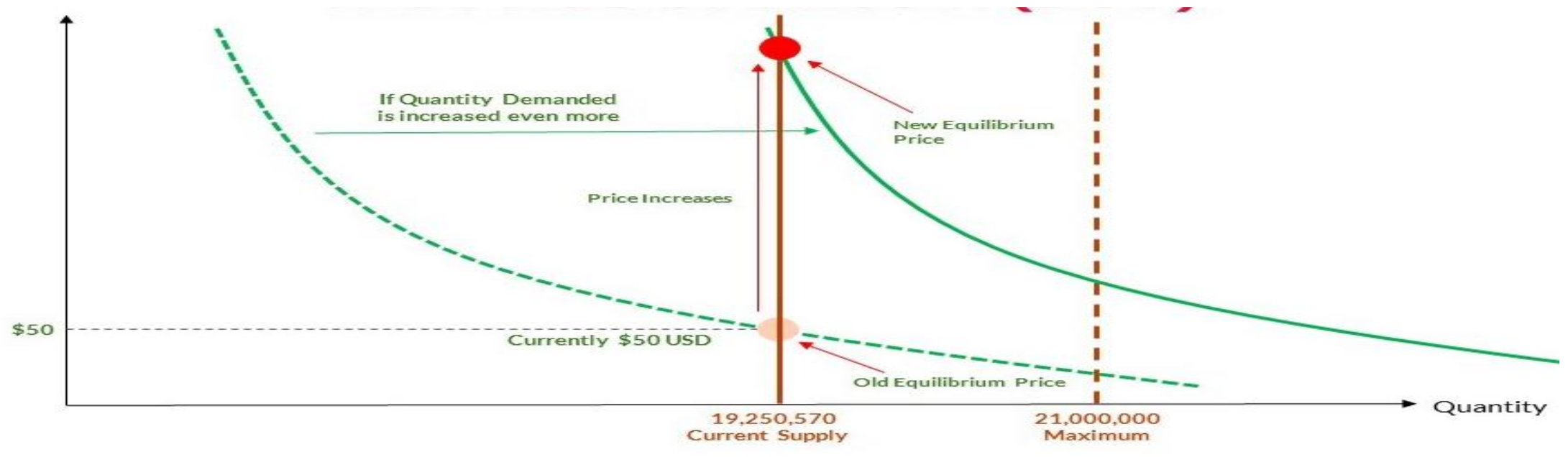

## ***NUMERICAL ANALYSIS BASED ON BITCOIN***

Let's, assume a numerical example for a better understanding, reference has been taken by (Ron and Shamir (2013)).

Let us assume a utility function of a buyer

$$U(x) = \log(x+b) - \log b$$

With b =0. We will pick following parameter values for the benchmark model. The length of a period is a day and the length of each trading session is 10 minutes (i.e., average block time). Setting $\beta$ = 0.9999 gives an annual discount factor of 0.97. The average Bitcoin supply in 2015 was 14342502.95. Consequently, the money growth rate per day in 2015 was $\mu = (1 + 25/14342502.95)^{6X24} = 1.00025$. This translates into an annual inflation rate of 9.6%. We use aggregate Bitcoin transactions to calibrate the rest of our parameters . We set $\sigma$ = 0.0178 to match the average fraction of Bitcoins spent per day, and set $\tau$ = 0.15596529 /1769.744292 = 0.000088129 to match the transaction fees data. The average transaction size is $\tau$ = 1769.744292/848.1232877 = 2.086659237. Finally, we use B = 6873428.441 which is the maximum number of average-sized transactions that the existing stock of Bitcoins. The distribution $F(\varepsilon)$ is set to capture the shape of the empirical distribution of transaction size reported in Ron and Shamir (2013). Based on this, we obtain an implied density function of the preference shocks $\varepsilon$ from our model and the confirmation lag N as optimally chosen in the transaction in our model gives a preference shock.

| | Per day | Per block |
|---|---|---|
| No of transactions | 122129.7534 | 848.1232877 |
| Estimated transaction volume (BTC) | 254843.1781 | 1769.744292 |
| Transaction fees (BTC) | 22.45900183 | 0.15596529 |

| | values | targets |
|---|---|---|
| $\beta$ | 0.999916553598325 | period length = 1 day |
| $\delta$ | 0.999999420487088 | $\delta = \beta^{1/(1+\bar{N})}$ |
| $\mu$ | 1.0003 | money growth rate |
| $\tau$ | 0.000088 | transaction fee |
| $B$ | 6873428 | max. no of average-sized transactions |
| $\sigma$ | 0.0178 | velocity per block (block length = 10 mins) |
| $\alpha$ | 1 | normalization |

# Preference Shock Distribution $f(\varepsilon)$

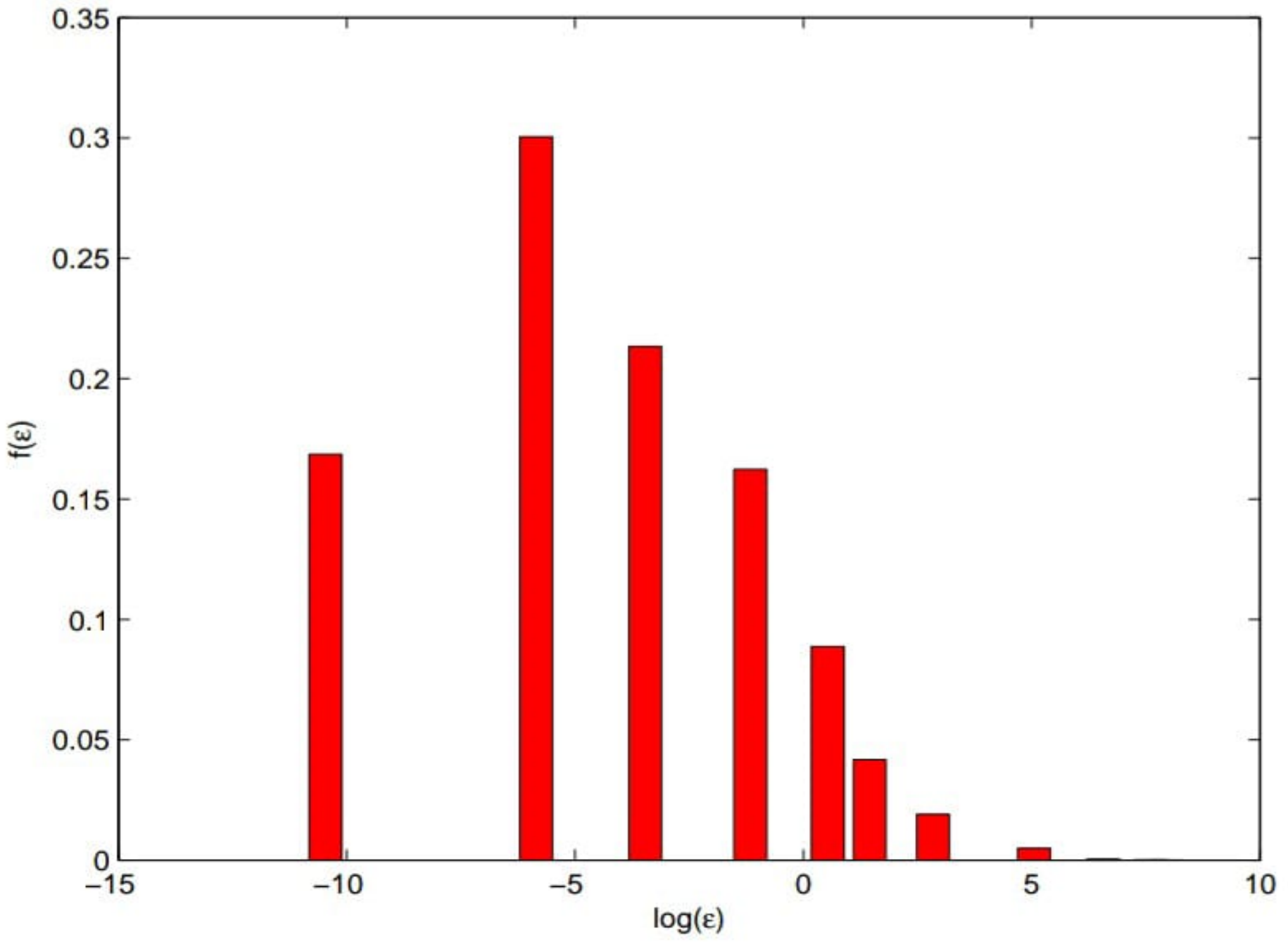

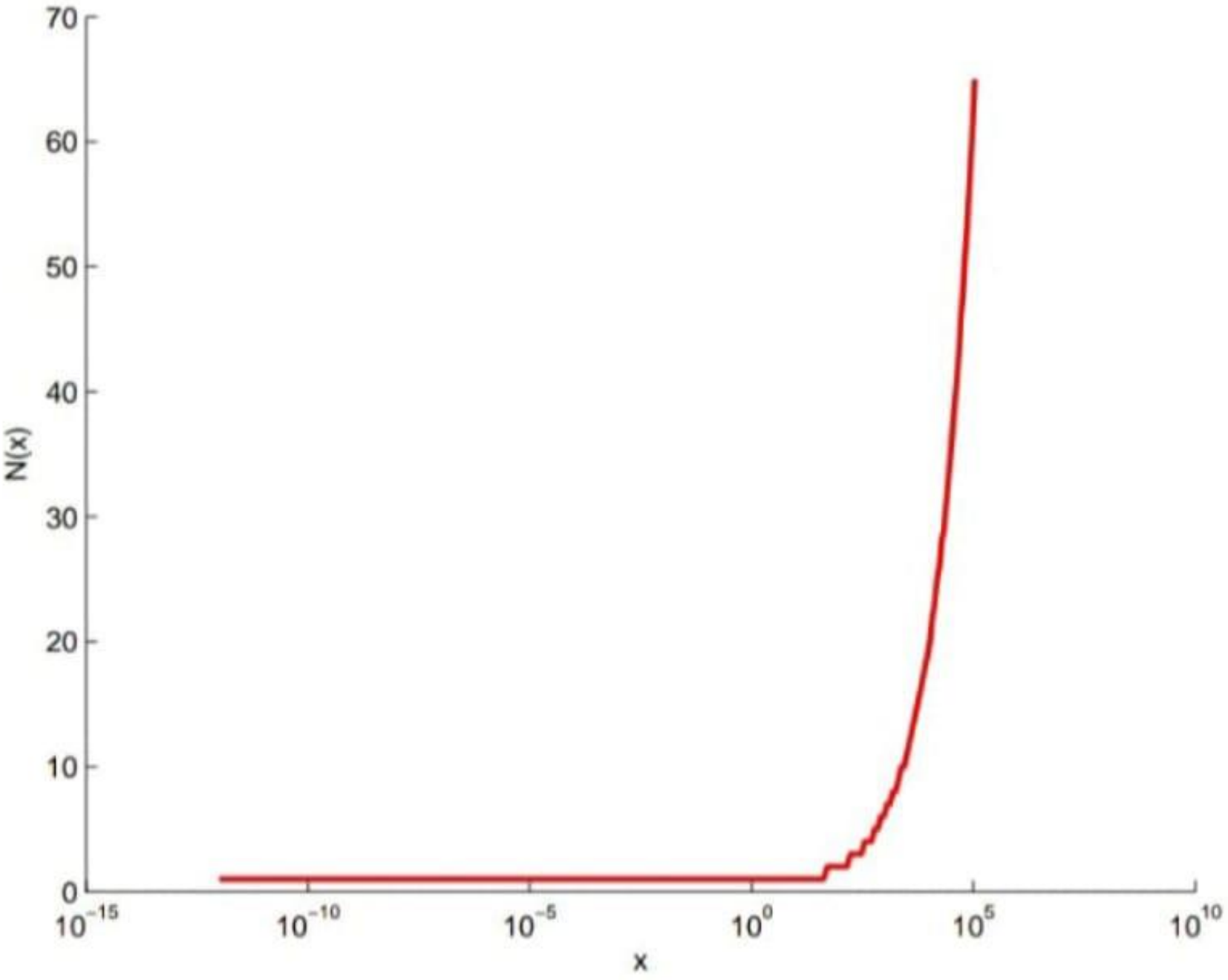

# ***INTRICACIES IN PRICE FLUCTUATIONS***

The price fluctuations of Bitcoin are influenced by a variety of intricate factors. Understanding these complexities can provide valuable insights for traders and investors. Here are some key intricacies in Bitcoin price fluctuations:

1. **Market Supply and Demand:**
   - **Limited Supply:** Bitcoin's supply is capped at 21 million coins, creating scarcity and potential for increased demand as more people seek to acquire it.
   - **Demand Variability:** Public interest, adoption, and media coverage can drive fluctuating levels of demand for Bitcoin, impacting its price.
2. **Regulatory Developments:**
   - **Legal Status and Regulations:** Changes in regulatory frameworks, government bans, or approvals significantly impact Bitcoin's price by affecting market participation and investor confidence and trust.
3. **Technological Factors:**
   - **Security Incidents:** Hacks, frauds, or security breaches can lead to a loss of confidence, resulting in immediate price drops. Protocol upgrades, forks, or changes in the underlying technology can influence price due to potential improvements, debates, or community disagreements.
4. **Market Liquidity and Exchange Dynamics:**

- **Liquidity Constraints:** Thin markets can result in more significant price swings due to limited liquidity and the potential for large buy or sell orders to impact the market.
- **Exchange Manipulation:** Market manipulation, wash trading, and spoofing on exchanges can distort price data and influence trader behaviour.

5. **Mining and Network Activity:**
   - **Hash Rate and Difficulty:** Changes in mining difficulty and hash rate can indicate changes in mining activity and network health, impacting the confidence in the network and, subsequently, the price.
6. **Market Speculation and Trading Behaviour:**
   - **Leverage and Derivatives:** The use of leverage and derivatives in trading can exacerbate price swings, leading to rapid gains or losses.

Understanding these intricacies and their interplay is important for anyone looking to engage in Bitcoin trading.

## CHALLENGES: HOW ATTACKERS AFFECT THE MARKET

The economics of this can be easily understood by calculating the scenario of an attack trying to generate an alternate blockchain faster than the honest chain. Even if this is accomplished, it does not throw the system open to arbitrary changes, such as creating value out of thin air or taking money that never belonged to the attacker. Nodes are not going to accept an invalid transaction as payment, and honest nodes will never accept a block containing them. An attacker can only try to change one of his own transactions to take back money he recently spent.

The race between the honest chain and an attacker chain can be characterized as a Binomial random Walk. The success event is the honest chain being extended by one block, increasing its lead by +1, and the failure event is the attacker's chain being extended by one block, reducing the gap by -1.

The probability of an attacker catching up from a given deficit is a Gambler’s Ruin problem. Suppose a gambler with unlimited credit starts at a deficit and plays potentially an infinite number of trials to try to reach breakeven. We can calculate the probability he ever breakeven, or that an attacker ever catches up with the honest chain:(source :Satoshi Nakamoto)

p = probability an honest node finds the next block

q = probability the attacker finds the next block

qz = probability the attacker will ever catch up from z blocks behind

$$q_z = \begin{cases} 1 & if\ p \le q \\ (q/p)^z & if\ p > q \end{cases}$$

For p > q, the probability drops exponentially as the number of blocks the attacker has to catch up with increases. With the odds against him, if he doesn't make a lucky lunge forward early on, his chances become vanishingly small as he falls further behind. We now consider how long the recipient of a new transaction needs to wait before being sufficiently certain the sender can't change the transaction. We assume the sender is an attacker who wants to make the recipient believe he paid him for a while, then switch it to pay back to himself after some time has passed. The receiver will be alerted when that happens, but the sender hopes it will be too late. The receiver generates a new key pair and gives the public key to the sender shortly before signing. This prevents the sender from preparing a chain of blocks ahead of time by working on it continuously until he is lucky enough to get far enough ahead, then executing the transaction at that moment. Once the transaction is sent, the dishonest sender starts working in secret on a parallel chain containing an alternate version of his transaction The recipient waits until the transaction has been added to a block and z blocks have been linked after it. He doesn't know the exact amount of progress the attacker has made, but assuming the honest blocks took the average expected time per block, the attacker's potential progress will be a Poisson distribution with expected value;

$$\lambda = z\frac{q}{p}$$

To get the probability the attacker could still catch up now, we multiply the Poisson density foreach amount of progress he could have made by the probability he could catch up from that point:

$$\sum_{k=0}^{\infty} \frac{\lambda^k e^{-\lambda}}{k!} \cdot \begin{cases} (q/p)^{(z-k)} & if\ k \le z \\ 1 & if\ k > z \end{cases}$$

## ***LORENZ CURVE – BITCOIN ANAYLSIS***

The Lorenz curve is a graphical representation of income or wealth distribution. It plots the cumulative share of total income or wealth held by the bottom x% of the population against the cumulative share of the population (y-axis). In a perfectly equal society, the Lorenz curve would be a straight line, known as the line of perfect equality.

However, Bitcoin doesn't have a clear income or wealth distribution like a traditional society. Ownership and distribution of Bitcoin are complex due to its decentralized nature and varying levels of participation and investment from individuals, institutions, miners, early adopters and many more . Therefore, creating a Lorenz curve for Bitcoin would not provide meaningful insights into income or wealth inequality.

## **GINI COEFFICIENT: APPLICATION WITH BITCOIN**

The Gini coefficient is a numerical measure of income or wealth inequality ranging from 0 (perfect equality) to 1 (maximum inequality). It's derived from the Lorenz curve. A Gini coefficient of 0 indicates that everyone has the same income, while a coefficient of 1 signifies that all income is held by one individual.

Again, applying the Gini coefficient directly to Bitcoin is challenging due to the decentralized and distributed nature of Bitcoin ownership. Different wallets, exchanges, and entities hold varying amounts of Bitcoin, making it difficult to measure income or wealth concentration accurately. while these concepts are valuable in analysing income or wealth distribution in traditional economic systems, they may not be directly applicable or insightful for analysing Bitcoin's economic structure and distribution. Alternative metrics and approaches specific to cryptocurrencies and blockchain technology may be more appropriate for assessing Bitcoin's economic aspects.

## ***BITCOIN AND ITS FUTURE***

The admits of G20 had a significant effect on cryptocurrency especially Bitcoin. G20 Finance Ministers and Central Bank Governors (FMCBGs), under India's leadership, have agreed to adopt the International Monetary Fund (MF) and Financial Stability Board (FSB)'s proposed plan for cryptocurrency (crypto-asset) regulations. Crypto-assets are a type of digital money that use special codes called cryptography to make sure that the transactions are secure and not duplicate or fake units. They are created by a network of computers and not issued or control by any central authority. International Monetary Fund (IMF) and Financial Stability Board (FSB) proposed a plan for regulating and supervising the crypto-assets acknowledging that banning them would not be an easy option. The plan is a policy framework and recommendation on understanding crypto assets to regulate for macroeconomic and financial stability.

The IMF-FSB report suggest banning of crypto could lead to: spillover risks, which means that crypto-asset activities and markets can move to other jurisdictions that have less strict or no regulations; costly enforcement, because crypto- assets are borderless and can be hidden or circumvented by using different technologies.

Additionally, it will hinder innovation and development. Hence, it's better to regulate it legally and financially for stability in investment, exchange, and consumer market

## ***CONCLUSION***

This paper has endeavoured to provide empirical evidence on the recent innovation Bitcoin, which can be considered as a new digital currency that is inextricably linked to a decentralized electronic payment system. However, in addition to its payment function, it also acts as an investment asset. This dual nature has proved crucial to its success so far and our paper has investigated both of these functions, ensuring that we paid sufficient attention to their interconnection. There are still many obstacles in the path ahead for Bitcoin, however. Perhaps the biggest is the legal status of the cryptocurrency, with some countries maintaining an outright ban, while others heavily restricting its use. it can be argued that the arrival of Bitcoin has ensured that a number of concepts, such as decentralized electronic transactions, distributed public ledgers, management of payment systems and the supply of the currency through cryptographic algorithms have been included in the science and practice of information systems and finance. However, it is also clear that the future of Bitcoin research will not be limited to any particular discipline. This research was limited to Bitcoin since it is the most used cryptocurrency having the highest market capitalization. Future research should also consider including other cryptocurrencies and using different data source. Lastly, cryptocurrency is a product of using cryptography to create a digital property. The frontier of digital property was popularized by the music industry’s shift to a cloud-based infrastructure. This frontier is still fairly new and unexplored, mainly populated by different types of media. Other forms of digital property may become as popular as music and cryptocurrency. Eight years ago, digital money was completely unheard of, and the creator of Bitcoin single handedly changed that. Cryptology, the root science beneath bitcoin and all cryptocurrencies, may be the mechanism behind the frontier for new and exciting digital inventions